# User Friendly Line CAPTCHAs

A. K. B. Karunathilake, B. M. D. Balasuriya and R. G. Ragel

Department of Computer Engineering, University of Peradeniya, Peradeniya 20400 Sri Lanka

Kazunn@gmail.com, bmdBalasuriya@gmail.com and roshanr@pdn.ac.lk

*Abstract-* **CAPTCHAs** or *reverse Turing tests* are real-time assessments used by programs (or computers) to tell humans and machines apart. This is achieved by assigning and assessing hard AI problems that could only be solved easily by human but not by machines. Applications of such assessments range from stopping spammers from automatically filling online forms to preventing hackers from performing dictionary attack. Today, the race between makers and breakers of CAPTCHAs is at a juncture, where the CAPTCHAs proposed are not even answerable by humans. We consider such CAPTCHAs as non user friendly. In this paper, we propose a novel technique for *reverse Turing test* - we call it the *Line CAPTCHAs* - that mainly focuses on user friendliness while not compromising the security aspect that is expected to be provided by such a system.

## I. INTRODUCTION

CAPTCHA is a challenge-response test used by computer programs to guarantee that the response is not generated by another computer (or program) [1]. The word CAPTCHA stands for Completely Automated Public Turing test to tell Computers and Humans Apart. CAPTCHAs are also known as reverse Turing test where a computer is the judge[1] or Human Interactive Proofs (HIP). CAPTCHAs are used to protect many types of websites, including free email providers, online ticket sellers, social networks, and blogs. For example, CAPTCHAs prevent ticket scalpers from using computer programs to buy large numbers of concert tickets, only to resell them at an inflated price. The main problem with almost all modern CAPTCHA methods is their user friendliness. Most of the unbroken CAPTCHA methods of today are difficult to understand and respond even by humans. Figure 1 depicts such a CAPTCHA method used by Yahoo! (http://www.yahoo.com) to prevent spammers filling their online forms.

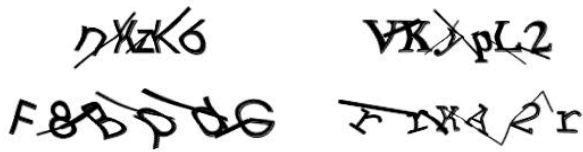

Fig. 1. An unbroken CAPTCHA used by Yahoo! (http://www.yahoo.com) to stop spammers filling their online forms. Human users are expected to read the distorted characters and fill them in a textbox correctly. This example depicts the difficulty of responding to such a CAPTCHA even by human.

---

[1] As opposed to the human judge in the *Turing test*

## II. APPLICATIONS OF CAPTCHAS

CAPTCHAs are becoming an important component of web security. Nowadays CAPTCHAs have a wide variety of applications on the web and we are discussing a number of them here.

### A. Online Polls

In November 1999, *slashdot.com* released an online poll asking, which was the best graduate school in computer science [2]. As is the case with most online polls, IP addresses of voters were recorded in order to prevent single users from voting more than once. However, students at Carnegie Mellon found a way to mark your ballot by using programs that voted for CMU thousands of times from computers around the world. CMU's score started growing rapidly. The following day, students at MIT wrote their own voting program and the poll became a contest between voting bots (programs). MIT finished with 21,156 votes, Carnegie Mellon with 21,032 and every other school (who didn't use computer programs to vote) with less than 1,000 votes. This incident brought the question of whether the result of any online poll can be trusted. The answer is no, unless the poll will guarantee that only humans can vote. A CAPTCHA can be used to make this guarantee.

### B. Free Email Services

Several companies (such as Google, Yahoo! and Microsoft) offer free email services, most of which suffer from a specific type of attack: bots that sign up for thousands of email accounts every minute [3]. This situation can be avoided or improved by requiring users to prove that they are human before they can get a free email account. Yahoo!, for instance, uses a CAPTCHA to prevent bots from registering for email accounts. Their CAPTCHA asks it user to read one or more distorted words such as the one shown in Figure 1 to guarantee that they are serving a human and not a computer bot. This particular challenge is a CAPTCHA because current computer programs are not as good as humans at reading distorted text. Therefore a computer bot will not be able to respond correctly to such a challenge – this type of a problem is also known as a hard AI (Artificial Intelligence) problem.

*C. Search Engine Bots*

Owners of some websites do not want their sites to be indexed by search engines. There is an HTML tag to prevent search engine bots from reading web pages. However, the tag does not guarantee that the bots will not read the pages; it only serves as to say no to bots used for indexing the web. Search engine bots, given that they usually belong to large esteemed companies, respect web pages that do not want to allow them in. However, these tags will not stop bots from indexing those web pages. In order to truly guarantee that bots will not enter a web site, CAPTCHAs are needed.

*D. Worms and Spam*

CAPTCHAs also offer a plausible solution against email worms and spam: email clients can adapt to a policy such as it only accepts an email if it knows there is a human behind the other computer from where that particular email is sent. A few companies, such as *http://www.spamarrest.com* are already marketing this idea.

*E. Preventing Dictionary Attacks.*

Pinkas and Sander [4] have suggested using CAPTCHAs to prevent dictionary attacks in password based systems. The idea is simple: prevent a computer from being able to iterate through the entire space of passwords by requiring a human to type the passwords.

### III. CHARACTERISTICS OF CAPTCHAS

There are several types of CAPTCHAs used in the recent past. People have used text images and sound CAPTCHAs. People can use any type of CAPTCHA method in their applications as long as the methods have the following characteristics [3]:

1. *Automation and grade-ability*: The test should be automatically generated and graded by a machine. This is the main requirement of a CAPTCHA. It is interesting to note that, although the test is both generated and graded by a machine, the machine will not be able to solve (or respond to) the test.

2. *Easy for human:* The test should be quickly and easily taken by a human user (the response time is typically within 30 seconds).

3. *Challenging and hard for machine:* The test should be based on a well-known hard AI problem and the best existing techniques should be far from solving the problem.
   An example problem that satisfies this requirement is "automatic image understanding" which is well known and has been investigated for more than three decades but is still without success. On the other hand, printed clean text OCR (Optical Character Recognition) is not a hard AI problem, as it can be solved with existing techniques.

4. *Universality:* The test should be independent of user's language, physical location, and education background. This guideline is motivated by practical considerations, and is especially important for companies with international customers such as Google, Yahoo! and Microsoft. It would be a difficult for Google if they had to localize a CAPTCHA test to 50 different languages. As an example, any digits-based audio CAPTCHA tests are not universal because there is no universal language on digits (even though visually they are the same). A different CAPTCHA test would have to be implemented for each different language, thus not cost effective. Strictly speaking, no CAPTCHA test can be absolutely universal, as there are no two humans who are the same in this world. However, we can make reasonable assumptions. For instance, it is reasonable to assume that a human who is using a computer knows the 10 digits and the 26 English alphabets.

5. *Resistance to no-effort attacks*: The test should survive no-effort attacks. No-effort attacks are the ones that can solve a CAPTCHA test without solving the hard AI problem.

6. *Robustness when database publicized:* The test should be difficult to attack even if the database, from which the test is generated, is publicized.

It is worth to note that these characteristics are a summary of what is presented in [3].

### IV. RELATED WORK

Even though CAPTCHA is a very new area of research, it has already attracted researchers from AI, cryptography, signal processing, and computer vision. The first idea related to CAPTCHA (*although it was not named CAPTCHA*) was written by Naor who wrote a note on this in 1996 [5]. The first working system was developed in 1997 by researchers at Alta Vista [6]. The goal of this system was to prevent automated programs from adding URLs to the search engine to disturb its search results. The specific technique they used was based on distorted characters, and it worked well in defeating regular optical character recognition (OCR) systems.

The term "CAPTCHA" was originally introduced in 2000 by Luis von Ahn, Manuel Blum, Nicholas J. Hopper (all of CMU), and John Langford (then of IBM) [7, 10]. The CMU team is one of the mostly active research team in CAPTCHA research from that point onwards. They have developed a number of concrete CAPTCHA systems [8] such as *Gimpy*, *Bongo* [9], *Pix*, *AnimalPix* and *reCAPTCHA* [11].

In the mean time, researchers at PARC and UC Berkeley have come up with a number of CAPTCHA techniques [12-14]. In their systems, they mainly explored the gap between human and programs in terms of reading poorly printed texts. In one of their systems, *Pessimal Print* [13], they reported close to zero recognition rates from three existing OCR systems: *Expervision*, *FineReader* and *IRIS Reader*. In another system, *BaffleText* [14], Chew and Baird further used non-English words to defend dictionary attacks.

In year 2000, Xu, Lipton and Ess, researchers at Georgia Tech have proposed a system to patch the security holes in E-commerce applications [15]. Another system proposed in [3] used distorted human faces to make CAPTCHAs.

Apart from the visual CAPTCHAs, there exist a number of audio CAPTCHAs. The general idea is to add noise and reverberation to clean speech such that existing speech recognizers can no longer recognize it. *Eco* [8], *Byan* [16] and the one presented in [18] are such systems.

The CAPTCHAs discussed up to now are all explicit CAPTCHAs. They are introduced as an additional exercise during browsing to prove that you are human. In 2005, Baird and Bentley proposed implicit CAPTCHAs [17]. An example from their list of such CAPTCHAs is an unconscious CAPTCHA in the form of a link on the text "*MORE PHOTOS*," rendered as an image. Such link text can be rendered deliberately to be difficult for machine readers, to form one in a series of effective implicit CAPTCHAs. Even though implicit CAPTCHAs eliminate the need for additional tasks to be performed by the users of explicit CAPTCHAs, they burden the website designers to come up with innovative CAPTCHAs to be embedded in their websites.

We realize that the race between CAPTCHA designers and breakers is a real one and the readers could refer to the following papers on the topic of breaking (and broken) CAPTCHAs: [19-23].

It is obvious that most of the CAPTCHAs with characters, images and audio are being broken. To improve the security, the developers increase the complexity of the system (by adapting to a harder AI problem) such as the character sequence presented in Figure 1. To solve such complex CAPTCHAs, sometimes even humans have to input values multiple times. Therefore, it is a big issue when considering the user friendliness of such CAPTCHAs. In addition we failed to find CAPTCHA methods which use lines and mouse movements for human interactive proofs. We consider mouse movements to be easily performed by human and therefore propose *Line CAPTCHA*, where lines and mouse movements are used to tell computer and human apart.

V. LINE CAPTCHAs

*A. Overview*

The CAPTCHA technique that we have proposed and implemented, *Line CAPTCHA*, basically gives users an image which has a distorted/broken line or a set of random lines. Then the user is expected to just drag the mouse along (while pressing the left button) the line continuously. When the user releases the left mouse button, the system performs the verification (also known as grading). At this moment we are proposing two *Line CAPTCHAs* (see Figure 2): (1) the user is expected to identify and draw on top of a line in a blurred image with a rich colourful background and (2) the user is expected to identify and draw on top a randomly segmented line (a continuous line that is segmented) again with a rich colourful background of line segments.

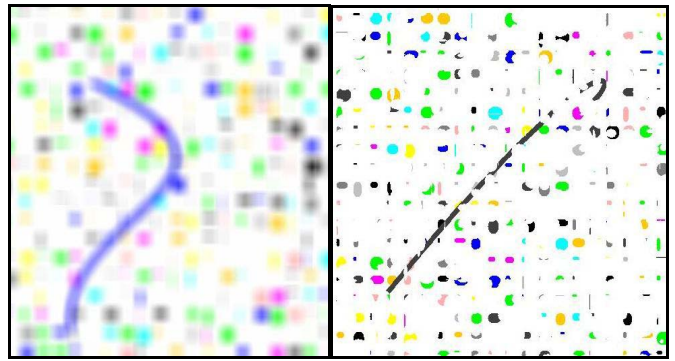

Fig. 2. Sample images of Line CAPTCHA. The one on the left is a blurred line on a colourful background and the one on the right is a randomly (pseudo random) segmented line.

In the first *Line CAPTCHA*, the assumption is the following: even though it is possible for a computer to identify and break such a CAPTCHA, the time taken to perform a line detection of a blurred line in an image with so many blurred other images is high such that an attack is not worth performing. However, this assumption could be challenged and therefore the validity of this *Line CAPTCHA*. However, the second *Line CAPTCHA* we have proposed has no such limitations. In the second *Line CAPTCHA*, we have a segmented line and a number of other line segments which are not part of any lines. Although, it is easy for a human to identify such a segmented line (see Figure 6), it is not the case for a computer. For a computer, there are many line segments and testing whether all these line segments will form a segmented line is tedious. This is the assumption made regarding the second *Line CAPTCHA* and the argument behind its validity as a CAPTCHA.

As mentioned earlier, one of the main objectives (if not "the" objective) of *Line CAPTCHA* is its user friendliness. A user who is expected to use a *Line CAPTCHA* will not even have to take his hand off the mouse. Just by using a "*mouse drag*" it is possible to respond to this CAPTCHA test. Random tests performed with users who have no familiarity with *Line CAPTCHAs* have shown more than 80% success rate in their first attempts. We will discuss the implementation steps of our *Line CAPTCHAs* in the following sub-section.

*B. Implementation*

Implementation of *Line CAPTCHA* is simple. The steps we followed in our implementation are described here. The implementation has two major phases and they are: (1) random image generation and (2) verification (or grading) of the user inputs. For the image generation, we followed three basic steps and they are: (1) background generation, (2) line generation and (3) distraction.

*1. Background Generation*

First the background of the image is generated. Random shapes with random colours are drawn in the background of the image as shown in Figure 3.

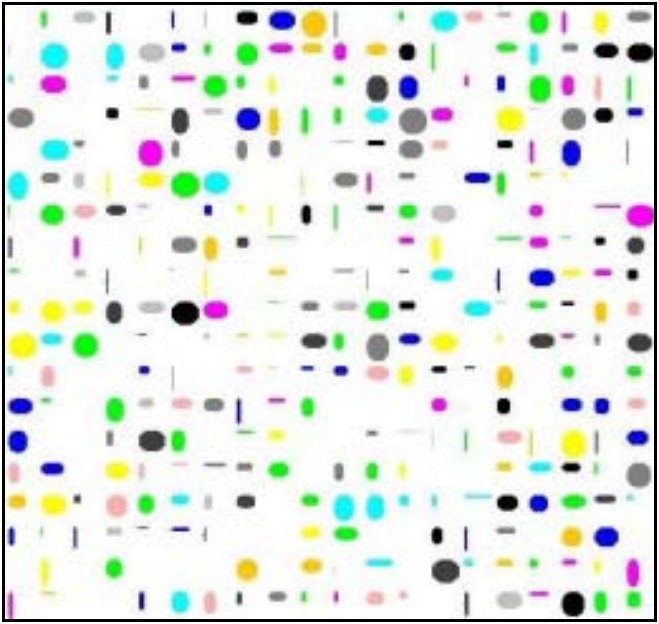

Fig. 3. The first step in image generation. Background of the image is generated with random shapes and colours. An image library of the implementation language is used for this purpose.

*2. Line Generation*

The second step is to draw the line on top of the background image. There could be several ways to draw lines. Line can be drawn using random coordinates or using mathematical equations. In our implementation, cubic curves were used to draw the line with necessary random points. These points are stored in the memory to perform the grading in the next phase of the CAPTCHA.

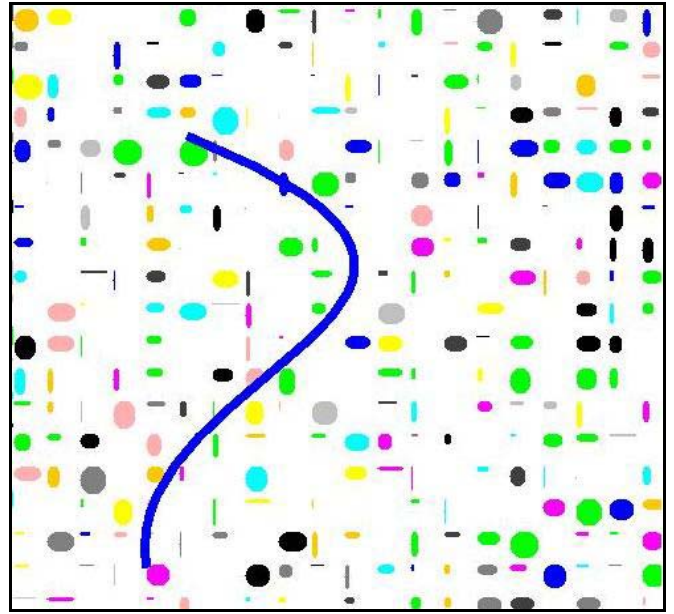

Fig. 4: The second step in image generation. The Line is drawn on the background.

*3. Distraction*

We call the next part of the image generation phase as the distraction phase. It should be noted that the image should not be distorted too much as the main goal of *Line CAPTCHA* is to increase the user friendliness. Too much of distraction will make it difficult for human. However, the drawn line should not be detected by a program as a single continuous line. As mentioned earlier in the overview, we propose two techniques first of which could be challenged. The two methods are blurring (as shown by an example in Figure 5) and line segmenting (as shown by an example in Figure 6).

The second phase of the implementation should be capable of grading human answers and verifying them (as correct or incorrect). This is achieved by storing the points used for drawing the original line in part two of phase one and then checking these points against the points generated by the mouse drag in this phase.

*Line CAPTCHAs* could be considered user-friendly than the existing CAPTCHA methods. In addition they could be used with any languages and therefore this technique is universal. However, this kind of CAPTCHAs cannot be used by visually impaired people. Broken line recognition (the second *Line CAPTCHA* proposed) is much harder to a computer program since the line is drawn randomly so it is difficult to use specific masks for recognition.

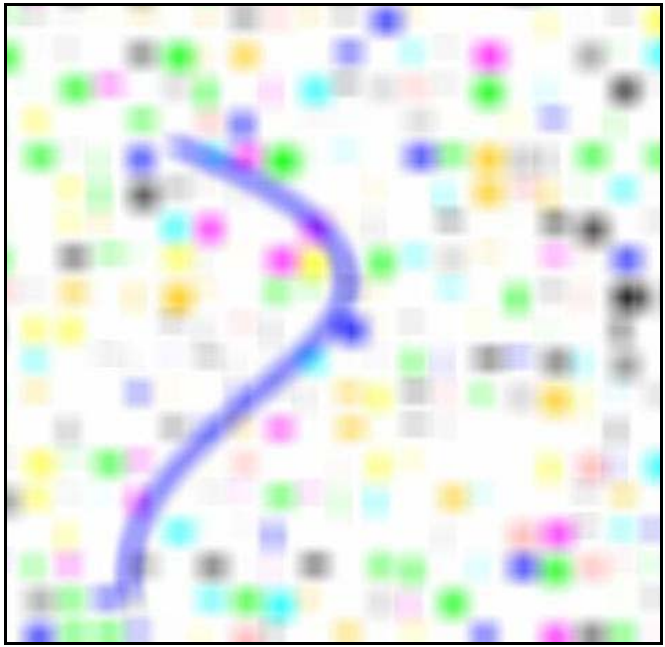

Fig. 5. The first type of *Line CAPTCHA* proposed. A blurred image with a blurred line in it. The user is expected to drag on top of the line.

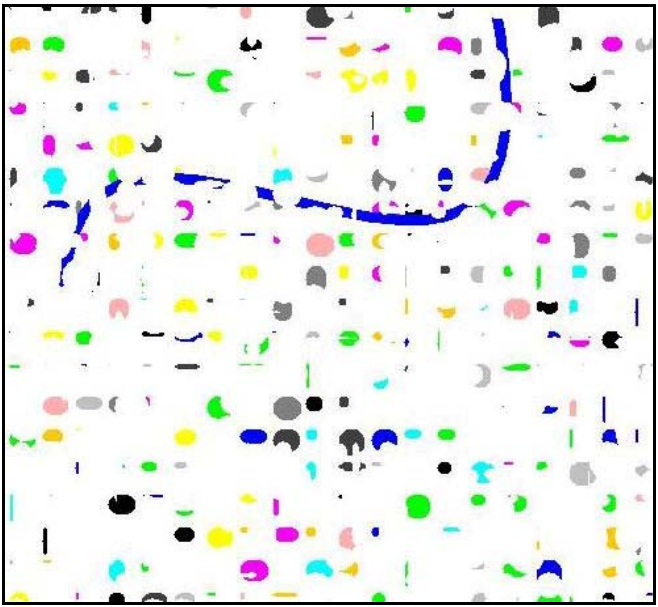

Fig. 6. The second type of *Line CAPTCHA* proposed. An image with a broken line (its done via line segmenting) and a large number of other objects, potential to be similar to the segments of the broken line.

The line segmenting technique could be subjected to colour separation attack - where an attacker separate pixels with different colours and then try to identify lines. To overcome this, we propose lines with multiple colour segments. A line with multiple colour segments could be used to overcome colour separation attack.

*C. More Line CAPTCHAs*

There could be more types of *Line CAPTCHAs* suggested and implemented with similar characteristics and user inputs. One such method is to use multiple lines with different colours and asking the user to select and draw on top a line with a particular colour as shown in Figure 7.

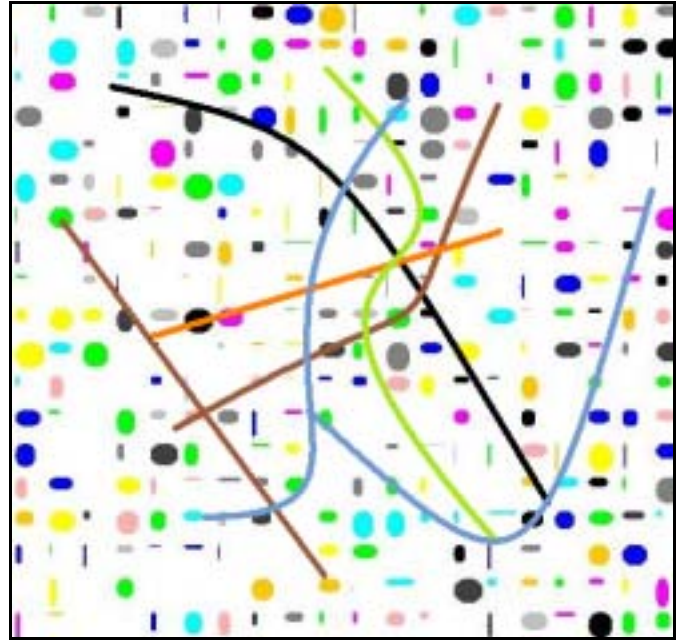

Fig. 7. An example for a different *Line CAPTCHA*. An image with multiple lines of different colours. The user could be asked to drag on top of a particular coloured line.

The validity of this *Line CAPTCHA* is in question for the same reason as the first CAPTCHA we proposed earlier. If a program can identify a line, then it can draw on top of a line randomly with a certain probability of success. In a secure CAPTCHA, the success rate for random inputs is considered to be less than 0.2%. So to get that rate there should be 500 lines in the image which is not practical. However, the reason for proposing this *Line CAPTCHA* is as follows: this *Line CAPTCHA* cannot be considered vulnerable for random input attack, since the computer has to first identify the lines before performing a random attack and this identification is going to consume significantly large time.

This *Line CAPTCHA* can be further enhanced for security by displaying the colour of the line to be drawn by using something like a reCAPTCHA [11]. reCAPTCHA, as shown in Figure 8, is a distorted text based CAPTCHA method that is considered unbreakable.

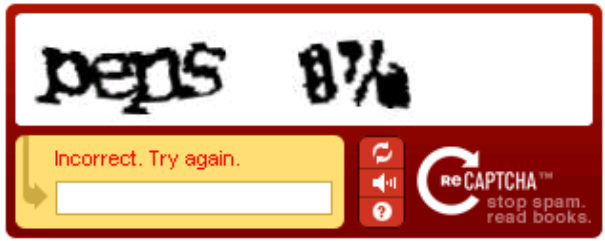

Fig. 8: A sample reCAPTCHA from http://recaptcha.net. It can be integrated with colour line based Line CAPTCHA for added security.

We consider reCAPTCHAs as non-user friendly, when it comes to reading and retyping every word (which is unknown to the users) tested. However, we believe that the user will be able to identify words for colours easily in reCAPTCHA (as the users are at an advantage of even predicting the word, since they expect a colour). Therefore this setup could still be considered user friendly and will have the added security.

## VI. CONCLUSION

CAPTCHAs or *reverse Turing tests* are used by programs (or machines) to differentiate humans and machines apart. The race between makers and breakers of CAPTCHAs is at a juncture where the CAPTCHAs proposed today are not answerable even by humans due to their complexity and non user friendliness. We consider CAPTCHAs that are unanswerable by human as non-user friendly. In this paper, we have proposed a novel CAPTCHA technique, known as *Line CAPTCHAs* which are distorted lines on non-plain background and are solved by mouse drags. *Line CAPTCHAs* mainly focus on user friendliness while not compromising the security that such systems are expected to provide.

## ACKNOWLEDGMENT

We would like to thank Dhammika Elkaduwe of Computer Engineering, University of Peradeniya for his valuable feedback on this project.

## REFERENCES

[1] Luis von Ahn, Manuel Blum, Nicholas J. Hopper, and John Langford, "CAPTCHA: Using Hard AI Problems for Security" Computer Science Dept., Carnegie Mellon University, Pittsburgh PA 15213, USA. IBM T.J. Watson Research Center, Yorktown Heights NY 10598, USA

[2] Ahn, L., Blum, M., and Hopper, N. J. (2002) Telling humans and computers apart (Automatically) or How lazy cryptographers do AI, *Technical Report CMU-CS-02-117*

[3] Yong Rui and Zicheng Liu, "ARTiFACIAL: Automated Reverse Turing test using FACIAL features," *Multimedia Systems*, Vol. 9, No. 6. (1 June 2004), pp. 493-502

[4] Benny Pinkas and Tomas Sander: Securing passwords against dictionary attacks. ACM Conference on Computer and Communications Security 2002: 161-170

[5] Naor, M. (1996) Verification of a human in the loop or identification via the Turing test, *unpublished notes*

[6] AltaVista's Add URL site: http://altavista.com/sites/addurl/newurl (1997)

[7] Wikipedia, http://en.wikipedia.org/wiki/CAPTCHA

[8] CAPTCHA website (2000) http://www.captcha.net

[9] Bongo, http://gs9.sp.cs.cmu.edu/cgi-bin/bongo

[10] Ahn, L., Blum, M., and Hopper, N. J. (2002) Telling humans and computers apart (Automatically) or How lazy cryptographers do AI, *Communications of the ACM*, Vol. 47, 2, February 2004, pp 57-60

[11] Luis von Ahn, Benjamin Maurer, Colin McMillen, David Abraham, Manuel Blum, reCAPTCHA: Human-Based Character Recognition via Web Security Measures, *Science*, Vol. 321, September 2008, pp. 1465-1468.

[12] Baird, H.S., and Popat, K. (2002) Human Interactive Proofs and Document Image Analysis, *Proc., 5th IAPR Workshop on Document Analysis Systems*, Princeton, NJ

[13] Chew, M. and Baird, H. S. (2003) BaffleText: a Human Interactive Proof," *Proc., 10th IS&T/SPIE Document Recognition & Retrieval Conf.*, Santa Clara, CA.

[14] Coates, A., Baird, H., and Fateman, R. (2001) Pessimal print: a reverse Turing test, *Proc. IAPR 6th Int'l Conf. on Document Analysis and Recognition*, Seattle, WA, pp. 1154-1158

[15] Xu, J., Lipton, R., and Essa, I. (2000) Hello, are you human?, *Technical Report (GIT-CC-00028)*

[16] Chen, N., Byan, http://drive.to/research

[17] Henry S. Baird and Jon L. Bentley, Implicit CAPTCHAs, *Proceedings, IS&T/SPIE Document Recognition & Retrieval XII Conference*, San Jose, CA, January 16-20, 2005

[18] Kochanski,G. et al. (2002): "A Reverse Turing Test using speech", *In: Proceedings of the Seventh International Conference on Spoken Language Processing (ICSLP2002 - INTERSPEECH 2002)*, Denver, Colorado, USA, September 16-20, 2002, 1357-1360

[19] Kumar Chellapilla, Patrice Y. Simard "Using Machine Learning to Break Visual Human Interaction Proofs (HIPs)" *Microsoft Research One Microsoft Way Redmond*, WA 98052

[20] Philippe Golle, Machine Learning Attacks Against the Asirra CAPTCHA *in the proceedings of CCS'08*, October 27–31, 2008, Alexandria, Virginia, USA, ACM 2008

[21] Greg Mori and Jitendra Malik, Recognizing Objects in Adversarial Clutter: Breaking a Visual CAPTCHA, *In Proceedings of the IEEE Computer Society Conference on Computer Vision and Pattern Recognition*, 2003

[22] Gabriel Moy, Nathan Jones, Curt Harkless, and Randall Potter, Distortion Estimation Techniques in Solving Visual CAPTCHAs, *In Proceedings of the IEEE Computer Society Conference on Computer Vision and Pattern Recognition*, 2004

[23] Jennifer Tam, Jiri Simsa, Sean Hyde and Luis Von Ahn, Breaking Audio CAPTCHAs, *Computer Science Department Carnegie Mellon University*, 2008